\def\breakon{\end{multicols}\widetext\vspace{-.2cm} 
\noindent\rule{.48\linewidth}{.3mm}\rule{.3mm}{.3cm}\vspace{.0cm}} 
\def\breakoff{\vspace{-.2cm} 
\noindent 
\rule{.52\linewidth}{.0mm}\rule[-.27cm]{.3mm}{.3cm}\rule{.48\linewidth}{.3mm} 
\vspace{-.3cm} 
\begin{multicols}{2} 
\narrowtext} 
\def\beq{\begin{equation}} 
\def\enq{\end{equation}} 
\def\bqa{\begin{eqnarray}} 
\def\eqa{\end{eqnarray}} 

\documentstyle[aps,prb,multicol,epsfig]{revtex} 

\begin{document} 
 
\draft 

\widetext 
 
\title{P-wave pairing and ferromagnetism in the  
metal-insulator transition in two dimensions} 
 
\author{Claudio Chamon$^a$, Eduardo R. Mucciolo$^b$, 
and Antonio H. Castro Neto$^a$\cite{riverside}}

\address{ $^a$ Department of Physics, Boston University, Boston, MA 
02215 \\ $^b$ Departamento de F\a'{\i}sica, Pontif\a'{\i}cia 
Universidade Cat\'olica do Rio de Janeiro, \\ Caixa Postal 38071, 
22452-970 Rio de Janeiro, Brazil} 
 
\date{\today} 
 
\maketitle 
 
  
\begin{abstract} 
  Based on recent experimental evidence for a spin polarized ground
  state in the insulating phase of the two-dimensional electron
  system, we propose that ferromagnetic spin fluctuations lead to an
  attractive interaction in the triplet channel and cause p-wave
  pairing in the conducting phase. We use the Landau Fermi liquid
  phenomenology to explain how the enhanced spin susceptibility near
  the critical density yields an attractive potential, in a similar
  mechanism to superfluidity in $^3$He. As the density is decreased,
  the p-wave order parameter undergoes a transition from a unitary to
  a nonunitary state, in which it coexists with ferromagnetism for a
  range of densities. As the density is further reduced, the pairing
  amplitude vanishes and the system is described by a ferromagnetic
  insulator. Thus, we find two quantum critical points as a function
  of density associated with the polarization of the paired state and
  ferromagnetism. We explain the magnetotransport measurements in
  parallel and perpendicular magnetic fields and propose a shot noise
  experiment to measure the pair charge.
\end{abstract} 
 
 
 
 
\begin{multicols}{2} 
 
\narrowtext 
 
\section{Introduction} 
\label{sec:Intro} 
 
It has become self-evident that the spin properties of interacting
electrons or holes in two-dimensions (2D) play a central role in the
transport properties of these systems, in particular in the possible
metal-to-insulator transition (MIT) observed in a number of different
material systems, such as Si-MOSFETs, n- and p-doped GaAs, AlAs, and
SiGe. \cite{RMP} Recent data in Si-MOSFETs by Shashkin {\it et al.}
\cite{Shashkin} and Vitkalov {\it et al.} \cite{Vitkalov} on the
saturation of the conductance as a function of the magnetic field
parallel to the 2D plane (see also Ref. \onlinecite{Pudalov1}),
combined with previous analysis of Shubnikov-de Haas oscillations in
tilted magnetic field,\cite{vitkalov2,Okamoto} suggest that the
insulating state is spin polarized.
 
These experimental observations revive an unresolved theoretical
problem on the possible phases of electronic systems in 2D as a
function of the interaction strength (or, alternatively, the density)
even in the idealized clean systems. Bergman and Rice
\cite{Bergman+Rice} raised the possibility that as the density is
decreased, there is a transition from a paramagnetic Fermi liquid into
a ferromagnetic Fermi liquid state. Quantum Monte Carlo studies by
Tanatar and Ceperley\cite{Tanatar+Ceperley} have considered three
different electronic states, a paramagnetic liquid, a ferromagnetic
liquid, and a Wigner crystal, and found a transition from the
paramagnetic liquid to the Wigner crystal at $r_s\approx 37$. However,
the energies of these three phases become rather close for a range of
$r_s$. Thus, it is not unreasonable that either improved energy
estimations, or disorder effects, may bring the energy of the
ferromagnetic state to lower values, so that it may exist for a window
of densities between the paramagnetic liquid and the Wigner
crystal. Indeed, perturbative Renormalization Group (RG) calculations
for disordered and interacting electrons in 2D by
Finkelstein\cite{Finkelstein} have pointed out a runaway flow in the
triplet channel even in the limit of low densities (see also
Ref. \onlinecite{Lee}). This runaway has been recently interpreted as
a tendency towards ferromagnetism in the diffusive (metallic)
regime,\cite{BK,CM} suggesting that disorder may also help trigger a
full spin polarization at low densities. Recent numerical studies have
also indicated a strong tendency towards spin polarization in the
localized regime.\cite{benenti}
 
\begin{figure} 
\hspace{-.2cm} 
\epsfxsize=8cm 
\vspace{.5cm} 
\epsfbox{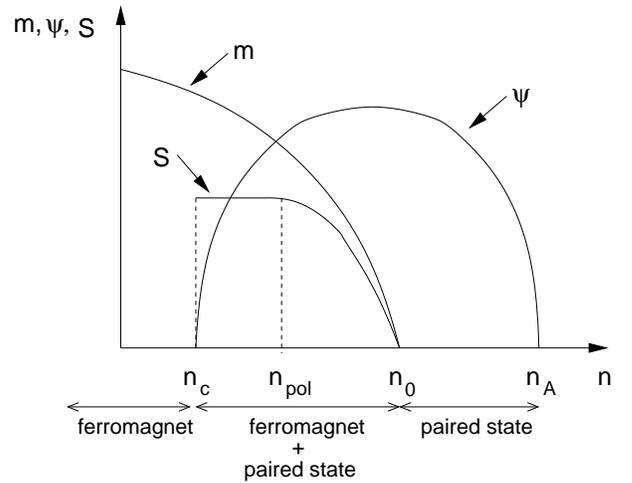} 
\caption{Variation of the relevant order parameters as a function of 
electron density (mean-field theory). $m$ is the magnetization, $\Psi$
is the p-wave pairing amplitude, and $S$ denotes the pair spin
projected along $m$. Spontaneous magnetization sets in at the critical
density $n_0$, while at $n_c$ the systems becomes a disordered
ferromagnetic insulator. $n_{\rm pol}$ is the density where the pairs
become fully spin polarized.}
\label{phase diagram} 
\end{figure}                                                      
 
Provided the recent experimental evidence from transport measurements
(not susceptibility measurements) for the ferromagnetic state of
dilute 2D electrons, we investigate the effect of enhanced spin
fluctuations on the paramagnetic side of the transition. We assume
that the paramagnetic state can be described, for densities above a
critical density $n_0$ and for finite temperatures below the Fermi
energy, by the Fermi liquid phenomenology. As one approaches the
critical density, one of the Landau Fermi liquid parameters, $F_0^a$,
which renormalizes the spin susceptibility, crosses the minimum bound
for the Pomeranchuk's stability condition.\cite{Pines&Nozieres} We
argue that the proximity to the ferromagnetic instability leads to an
attractive interaction for a range of densities preceding the
ferromagnetic transition. The attraction, in the triplet channel,
leads to p-wave pairing. Once the system enters the ferromagnetic
state, the paramagnon exchange mechanism for the attraction rapidly
decreases, and these two phases compete to the point where the p-wave
state ceases to exist at a density $n_c$. Therefore, there are two
quantum critical points in the phase diagram at densities $n_c$ and
$n_0$ (likely very close), and there is an intermediate region in
densities where the two phases coexist, but in which the p-wave
pairing is in a nonunitary state. At the mean field level, the 
phase diagram is shown in Fig. \ref{phase diagram}.
 
The possibility of unconventional pairing in an electronic system has
been recently considered in the context of layered Sr$_2$RuO$_4$. In
the Sr$_{n+1}$Ru$_n$O$_{3n+1}$ series, where $n$ determines the number
of RuO$_2$ planes in the unit cell, ferromagnetic states are observed
for $n>3$. It is believed that the proximity to ferromagnetism plays
an important role for superconductivity in the Ruthenates, and p-wave
symmetry was proposed on the basis of similarities to
$^3$He.\cite{manfred,baskaran} Although a direct transition between
ferromagnetic and superconducting states has not been experimentally
observed, the Ruthenates show similarities to the problem studied in
this work. We should point out, however, that in Ruthenates there is
true 3D long range order, while we discuss systems that are truly
two-dimensional and therefore subject to strong fluctuations.
 
We also draw a strong analogy between the 2D electronic states and the
3D $^3$He systems in a number of ways.\cite{spivak} For one, we argue
that the Fermi liquid phenomenology should not be dismissed in
describing the 2D interacting electronic system in the metallic side
of the transition at finite temperatures. One of the usual concerns
that is raised against the Fermi liquid state in the 2D problem at low
densities is that the ratio between the Coulomb and kinetic energies
is about a factor of ten. This logic can be misleading, a) since the
contributions from exchange and correlations reduce this ratio, and b)
Fermi liquid parameters, which measure the strength of the interaction
relative to the kinetic energy,\cite{Pines&Nozieres} are typically
large compared to unity even for 3D $^3$He. For example, the Landau
parameters for $^3$He at high pressure ($27$ bar) that renormalize the
compressibility, magnetic susceptibility, effective mass, and spin
precession rate are $F_0^s=68.17$, $F_0^a=-0.76$, $F_1^s=12.79$, and
$F_1^a=-1.00$.\cite{baym} Even though most Landau parameters are very
large when compared to unit, the normal phase of $^3$ He is very well
described by Landau's Fermi liquid theory. Indeed, if any, the real
question is why Landau's phenomenology works so well, way beyond the
perturbative regime where RG arguments for fermions \cite{Shankar}
justify the stability of the Fermi liquid.
 
We show that we can consistently interpret the recent data on the 
2D MIT transition close to the critical density 
\cite{Shashkin,Vitkalov,Pudalov1} as due to the enhancement of the Landau 
parameter $F_0^a$. The paramagnon exchange mechanism can be
responsible for an attraction in the triplet channel in the 2D
electron problem, in complete analogy with the problem of 3D
$^3$He. The major difference between these two systems is the
dimensionality: in 2D true superconductivity or superfluidity is only
possible at zero temperature. Strong fluctuations in 2D do not allow
for long range order. For a singlet paired state in 2D with U(1)
symmetry algebraic order can be established below the
Kosterlitz-Thouless (KT) temperature.\cite{kt} However, for a triplet
paired state where the order parameter is a complex vector and the
symmetry group is nonabelian, superconductivity can only be
established at $T=0$.\cite{footKT} We thus propose that the insulating
state is ferromagnetic, while the metallic state corresponds to a
paired p-wave state. The possibility of {\it singlet}
superconductivity in the observed conducting phase of the 2D electron
systems was suggested by Phillips {\it et al.}\cite{philip} and Belitz
and Kirkpatrick;\cite{belitz} here we present a possible mechanism for
pairing (without finite temperature long-range order) in the {\it
triplet} channel.\cite{philipP}
 
We would like to stress that we do not address in this paper the 
reason why the measured conductance in Si-MOSFETs seems to saturate in 
the triplet paired state when $T \to 0$. In the case of singlet paired 
superconducting 2D materials (thin films) and 2D Josephson junction 
arrays, where a finite KT transition should be observed together with 
the vanishing of the resistivity, a saturation of the conductivity is 
also observed.\cite{dissipative} The source and precise mechanism for 
these dissipative effects are presently unknown, despite some recent 
theoretical efforts.\cite{Sudip+SteveK,Phillips2} 
 
There is a natural question regarding the possibility of a paired
state in the 2D electronic systems where the MIT is observed: Why can
one have pairing if the conductance is of order $e^2/h$ near the
transition? Naively, if one uses intuition from noninteracting
electrons, then the bare value of the conductance is $G_0 = (2e^2/h)
(k_F \ell)$; weak localization corrections, perturbative in $(k_F
\ell)^{-1}$, are added to this bare value. This would imply that $k_F
\ell \approx 1$ at the transition, and therefore disorder is too
strong and pair breaking. However, the value of $\ell$ that one reads
from this naive argument is {\it not} a measure of disorder alone. The
energy scale of the interactions is larger than the Fermi energy of
the 2D electron systems near the transition, thus the scattering of
electrons even if the disorder is weak should be large and
dominant. Therefore, a dimensionless conductance of order unity does
not necessarily imply strong disorder. For example, it is known that
near 2D superconducting-insulator (SC-I) transitions the dimensionless
conductance is of order one even in the absence of disorder.\cite{bose
hubbard} We claim that the same happens in the context of the MIT
transition discussed here. We also give arguments showing, based on
Fermi liquid theory, that near the ferromagnetic transition the
paramagnetic scattering can provide for conductances of order unity.
 
The paper is organized as follows: In Section \ref{sec:Landau} we
discuss the Landau's Fermi liquid phenomenology applied to the
paramagnetic phase of the 2D electronic system, and the Pomeranchuk
instability leading to ferromagnetism at low densities. We argue that
the enhancement in spin fluctuations due to the proximity to the
ferromagnetic state causes an attractive interaction in the triplet
channel and p-wave pairing. Section \ref{sec:two-fluid} contains a
discussion of a two-component model for the MIT transition, where we
argue that, due to electron-electron interactions in the presence of
disorder, the dimensionless conductance due to paramagnetic scattering
can become of order unit close to the transition. The mean-field phase
diagram of the problem is established in Section \ref{sec:one} using a
generic Ginzburg-Landau free energy for a p-wave paired state coupled
to a ferromagnetic order parameter. In Section \ref{sec:experiments}
we compare our results to the available experimental data for parallel
and perpendicular magnetic fields for the 2D electron gas in a
Si-MOSFET and propose new shot noise experiments that can test our
theory. Section \ref{sec:conclusions} contains our conclusions.
 
 
\section{The P-Wave Paired Metallic Phase} 
\label{sec:Landau} 
 
Our starting point is a Landau Fermi liquid theory for the metallic
phase of the 2D electronic system. We consider first the effects of
interactions, and then those of disorder. Let us briefly review and
then apply the Landau phenomenology to the experimental observations
on Si-MOSFETs. Consider an isotropic Fermi liquid with planar density
$n$, Fermi momentum $k_F = \sqrt{2 \pi n/g_v}$, and Fermi energy $ E_F
= \hbar^2 k_F^2/2 m^\ast = \pi \hbar^2 n/g_v m^\ast$, where $m^\ast$
is the effective mass and $g_v$ accounts for the valley
degeneracy. The ground state of the problem is described in terms of
quasiparticles that fill up a Fermi sea up to the Fermi energy. The
change in the energy of a Fermi liquid due to changes in the
quasiparticle charge density, $\delta n(\vec k)$, and spin density
$\delta \vec\sigma(\vec k)$ is given by
\cite{Pines&Nozieres,baym}
\bqa 
 \delta E & = & \int d^2 k \, \epsilon_{\vec k}\, \delta n(\vec k) + 
 \int d^2 k\, d^2k^\prime\, f^s(\vec k,\vec k^\prime)\, \delta n(\vec 
 k)\,\delta n(\vec k^\prime) \nonumber \\ & & + \int d^2 k\, 
 d^2k^\prime\, f^a(\vec k,\vec k^\prime)\, \delta \vec\sigma(\vec k) 
 \cdot \delta \vec\sigma(\vec k^\prime), 
\eqa 
where $\epsilon_{\vec k}$ is the bare dispersion and $f^s(\vec k,\vec
k^\prime)$ and $ f^a(\vec k,\vec k^\prime)$ are the symmetric and
anti-symmetric Landau parameters, respectively. In 2D these parameters
can be expanded as
\bqa 
f^{a,s}_{\vec k,\vec k'} = \sum_{n=-\infty}^{+\infty}  
f^{s,a}_n e^{i n \theta_{\vec k,\vec k'}} 
\eqa 
where $n$ gives the angular momentum in the plane, and $ \theta_{\vec 
k,\vec k^\prime}$ is the angle between $\vec k$ and $\vec 
k^\prime$. It is useful to define dimensionless parameters $F_n^{a,s} 
\equiv N(0) f^{s,a}_n$, where $N(0) = g_v m^\ast/\pi\hbar^2$ is the 2D 
density of states at the Fermi energy.
 
The stability of the Fermi liquid state (or the Fermi surface) is
given, in Landau's theory, by the Pomeranchuk
criterion,\cite{Pines&Nozieres,baym} which in 2D can be written as
\bqa 
F^{s,a}_n > -1 
\label{pomeranchuk} 
\eqa 
for all values of $n$. Since all the physical quantities in Landau's
theory can be written in terms of the Landau parameters, a violation
of the Pomeranchuk criterion implies an instability of a physical
observable. The compressibility, for instance, is given by
\bqa 
\kappa = \frac{N(0)}{1+F^s_0} 
\eqa 
and an instability to phase separation implies that
$F^s_0<-1$.\cite{phase separation} In the same theory the effective
mass is given by
\bqa 
\frac{m^\ast}{m_b} = 1 + F^s_1, 
\label{effective mass} 
\eqa 
vanishing when $F^s_1=-1$ (here $m_b$ is the carrier band mass). The
magnetic susceptibility can be written as
\bqa 
\chi = \left(\frac{g_0 \mu_B}{2}\right)^2 \frac{N(0)}{1+F_0^a}, 
\label{sus} 
\eqa 
where $g_0 \approx 2$ is the bare (band) Land\'e $g$-factor. Thus, for
$F_0^a \approx -1$ the magnetic susceptibility diverges, indicating an
instability towards a magnetically ordered phase.
 
As one of the Landau parameters approaches the critical value given by
Pomeranchuk's criterion, there is a strong enhancement of the
interactions in the Fermi liquid. Consider, for instance, the case of
density-density interactions that are determined by $F_0^s$. It can be
shown that the induced density-density interaction in a Fermi liquid
(in the static limit) is given by the usual RPA expression
\cite{leggett} 
\bqa 
U_{\rho-\rho'} = \frac{1}{N(0)} \frac{F_0^s}{1+F_0^s}. 
\eqa 
Thus, close to the Pomeranchuk's instability $F_0^s \approx -1$ the 
interaction is very large and attractive, leading to phase 
separation. On the other hand, the induced spin-spin interactions in 
the same system is given by 
\bqa 
U_{\sigma-\sigma'} = \frac{1}{N(0)} \frac{F_0^a}{1+F_0^a}  
\, \, \vec \sigma \cdot \vec \sigma^\prime, 
\eqa 
where $\sigma$ is the electron spin. Therefore, when $F_0^a \approx 
-1$, that is, close to the magnetic instability, this interaction is 
also large and attractive, leading to pairing in a spin triplet 
channel. In fact, we can estimate the size of the pairing amplitude 
using the weak coupling BCS expression 
\bqa 
|\Delta_p| \approx E_F \, \, e^{-\left|\frac{1+F_0^a}{F^a_0}\right|}, 
\label{pairing amplitude} 
\eqa 
where the Fermi energy $E_F$ works as a cutoff in the problem because 
is the only energy scale present.\cite{leggett} Obviously, we have to 
consider Eq. (\ref{pairing amplitude}) carefully since, as the system 
approaches the instability, the attraction is very strong and the weak 
coupling expression breaks down; in this case, one should use a strong 
coupling approximation.\cite{Eliashberg} Thus, the expression in 
Eq. (\ref{pairing amplitude}) can still be used when the attraction is 
weak $|F_0^a|\ll 1 $ and is only a crude estimate when $F_0^a 
\approx -1$. 
 
In general, one expects the Landau parameters to be dependent on the 
electronic density $n$. Let us consider the situation of a Fermi 
liquid close to a magnetic instability that happens at $n=n_0$. The 
Landau parameter, $F_0^a(\delta)$, can be expanded close to the 
transition as 
\bqa 
F_0^a(\delta) = -1 + \alpha \delta + {\cal O}[\delta^2], 
\label{landau expansion} 
\eqa 
where $\alpha>0$ is a constant and  
\bqa 
\delta = \frac{n-n_0}{n_0} 
\eqa 
measures the distance from the quantum critical point. In
Eq. (\ref{landau expansion}) we disregard higher order terms in the
density variations around the critical point. Observe that in this
case the magnetic susceptibility can be written from Eq. (\ref{sus})
as
\bqa 
\chi(\delta) &\approx& \frac{(g \mu_B)^2 N(0)}{4 \alpha\delta}, 
\label{divergences} 
\eqa 
showing that the susceptibility diverges linearly with the distance 
from the critical point. Notice, from Eq. (\ref{pairing amplitude}), 
that the weak coupling expression for the pairing amplitude close to 
the transition is given by 
\bqa 
|\Delta_p(\delta)| \approx E_F e^{- \alpha |\delta|} \approx E_F. 
\label{delta ef} 
\eqa 
This result indicates that the pairing amplitude is of order of the
Fermi energy in the system. In the case of the 2D Si-MOSFETs, $E_F$ is
usually of order of $5$ K because of the low electronic density. The
critical temperature, $T_c$, however, remains zero because of the
dimensionality of the system, otherwise this problem would be a case
of high temperature superconductivity. Notice that the number of pairs
is $n_p = n/2$ at zero temperature.
 
Away from the transition we can classify the behavior of the system
depending on the full density dependence of $F_0^a$. At large enough
densities, $n>n_A$, $F_0^a$ should become positive because of the
screening of the electronic interactions by backflow effects. In this
case, without the help of pairing, the localization effects should
dominate, and the system should become an Anderson insulator. So it is
only in this high density ($n>n_A$) regime that one could possibly
attempt to apply ideas established through the scaling theory of
localization\cite{Gang4} for noninteracting electrons. For $n_0<n<n_A$
we have $-1<F_0^a<0$ and pairing is effective in delocalizing the
electrons, leading to the metallic state observed experimentally. In
this regime, naive intuition based on noninteracting electrons should
not apply. Finally, for $n<n_0$, the system becomes a ferromagnet. We
will return to this regime later when we consider a Landau-Ginzburg
theory describing the system. We show that there is a coexistence
region with both p-wave pairing and ferromagnetism for densities
$n_c<n<n_0$, and for $n<n_c$ the physics of the problem is the one of
a ferromagnetic insulator.

\subsection{The Effect of Disorder in the Paired State} 
 
We have argued that the proximity to the ferromagnetic instability
induces pairing in the Fermi liquid state. The effects of strong
interactions are built in in this picture, but we still need to
discuss the effect of disorder in this paired state (which we argue in
Section \ref{sec:one} should be in fact a fully gapped $p_x \pm ip_y$
state). The question to be addressed is: will the paired state survive
disorder? As it is known for the case of strongly coupled
superconductors, which is the case here, when the electron mean free
path, $\ell$, becomes of the order of the coherence length, $\xi_{\rm
sc}$, pairing is suppressed.\cite{abrikosov} Since the pairing
amplitude is essentially of order of $E_F$ [see Eq. (\ref{delta ef})],
the coherence length is of the order of the Fermi wavelength,
\bqa 
\xi_{\rm sc} = \frac{v_F}{|\Delta_p|} \approx \frac{1}{k_F}, 
\label{xis} 
\eqa 
and of the size of the interparticle spacing. Therefore, pairing
should survive as long as $\ell \gg \xi_{\rm sc}$, or equivalently,
$k_F \ell \gg 1$. The main question here is whether this condition is
satisfied in the heterostrutures where the MIT is observed. We argue
below in favor of this case.
 
The common belief is that close to the MIT transition $k_F \ell 
\approx 1$ because the dimensionless conductance at the transition is 
of order one. This indeed would be the case if the value of the
conductance would be completely fixed by the amount of disorder in the
sample. However, as we have shown, there are strong electron-electron
interactions close to the ferromagnetic transition (indeed
interactions are the {\it reason} for the instability, since $F_0^a
\rightarrow -1$). These interactions in the presence of localized
electronic spins, as we are going to show in Section
\ref{sec:two-fluid}, can provide a large contribution to the
resistance. In other words, the value of the critical resistivity,
$\rho_c$, can be of order $h/e^2$ from interaction effects even if
disorder is small. The concept of using $(k_F \ell)^{-1}$ as a measure
disorder is only good if one could determine what $\ell$ is
independently of a conductivity measurement, because otherwise strong
interaction effects will blend in and make such discussion
useless. Therefore, in strongly correlated systems one should be
careful in extending arguments that are only valid for noninteracting
systems.
 
>From a theoretical perspective, it is known that interactions alone
can lead to a universal conductivity of order $\sigma_0 = 4 e^2/h$. It
has been shown \cite{bose hubbard} that, for the 2D
superconductor-to-insulator transition in the clean Bose-Hubbard model
(belonging to the 3D-XY model universality class), the conductivity is
given by $\sigma \approx 0.285 \sigma_0$. It was argued in
Ref. \onlinecite{bose hubbard} that this transition is in the same
universality class of a superfluid-insulator transition for bosons
moving in a random potential. In $1+1$ dimensions, it is known that
the universality class of a superconductor to insulator transition for
a model of fermions with attractive interactions is the same as the
insulator to superfluid transition in a model of repulsively
interacting bosons.\cite{giamarchi-schulz} Since $1+1$ is the lower
critical dimension for this type of transition,\cite{bose hubbard} we
expect this type of result to hold in our case as well. Indeed, there
is strong experimental evidence for these results in the
superconducting to insulator transition in amorphous superconducting
thin films.\cite{dissipative}

\section{The Two-Component Model} 
\label{sec:two-fluid} 
 
\begin{figure} 
\hspace{-.2cm} 
\epsfxsize=8cm 
\vspace{.5cm} 
\epsfbox{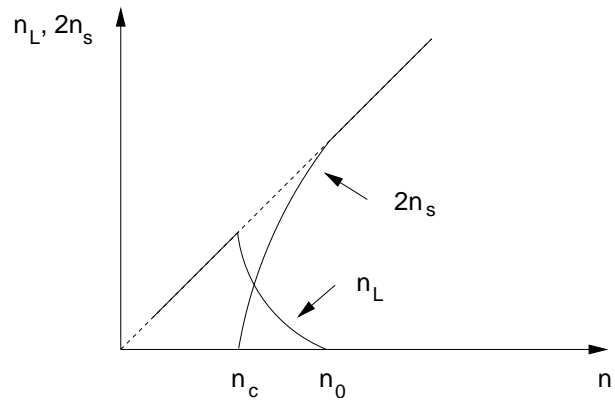} 
\caption{Densities of the localized, $n_L$, and itinerant, $2 n_p$, 
components as a function of total density near the critical region.}
\label{fig:twofluid} 
\end{figure}

Based on experimental evidence that the insulating phase in the 2D
electron system is a ferromagnetic state, we have argued in Section
\ref{sec:Landau} that the proximity to the spin polarized state
induces p-wave pairing on the metallic side of the MIT. In this
Section we study the transition between the p-wave paired state
and the ferromagnetic state using a two-component model. Let us split
the total electron density at $T=0$ into a localized ferromagnetic
component, $n_L$, and an itinerant, paired component, $n_p$:
\beq 
n=n_L + 2 n_p. 
\enq 
The dependence of both $n_L$ and $n_p$ as a function of $n$ is
represented schematically in Fig. \ref{fig:twofluid}. For densities
$n_0 <n <n_A$, all electrons are paired ($n_p=n/2$) at $T=0$, while
for very low densities, all electrons are in the insulating
ferromagnetic state ($n=n_L$). There is a region where the two
components coexist: as the system starts to spin polarize below $n_0$,
the density of pairs starts to decrease, vanishing at some critical
density $n_c<n_0$. At very low densities ($n \ll n_c$) the Wigner
crystal or Bragg glass phase becomes the ground
state.\cite{goma,vclong} In this work, however, we do not attempt to
study such phase, since it requires unscreened long range interactions
that are not included in our formalism. The study of the Wigner
crystal would then need a different starting point, far from the
MIT. For this reason, the Wigner crystal is absent in the phase
diagram of Fig. \ref{phase diagram}.

The appearance of a coexistence region is natural in the way we divide
the two densities. It also follows from the competitive nature of the
two phases, as the mechanism that leads to pairing - paramagnon
exchange - is strongly suppressed as the system spin polarizes. The
coexistence phase is also found within the Landau-Ginzburg theory of
Section \ref{sec:one}, as a direct consequence of having two competing
orders. The existence of metallic regions surrounding localized
puddles of charge has been confirmed experimentally through ingenious
local compressibility measurements in GaAs/AlGaAs
samples.\cite{weizmann} These experiments reveal that the fragmented
localized regions increase in number as the density is lowered towards
the MIT. The appearance of these fragments in the metallic phase
($n>n_c,n_0$) is consistent with the idea that some incipient
magnetism is always present in diffusive systems, even when the
interaction in the triplet channel is weak.\cite{CM} In regions where
strong magnetization occurs due to fluctuations in the wave functions,
the paramagnon mechanism is absent, thus suppressing pairing and
leading to localization. This is not taken into account in our
mean-field (homogeneous) treatment of the problem; had it been
considered, it would likely prolongate the $n_L$ curve towards
densities larger than $n_0$ (not shown in Fig. \ref{fig:twofluid}).

Next, we use this two-component picture to provide an argument on why
the critical conductivity near the MIT is of order $e^2/h$. Consider
the situation at finite temperature within the coexistence window
$n_{c}<n<n_0$. While at zero temperature all the delocalized electrons
(with density $2 n_p$) participate in the pairing, at finite
temperatures only a small fraction is really paired. The unpaired
state can be described by the Fermi liquid theory of Section
\ref{sec:Landau}. The localized ferromagnetic component is a source of
spin scattering for the itinerant Fermi liquid. The coupling
\beq 
\delta H = J\; \int d^2r\; \vec S_{L}(r) \cdot \vec S_{\rm it}(r) 
\enq 
follows from splitting the electrons into two components in the
effective Hamiltonian for the interacting electron liquid. Here
$S_L({\bf r})$ and $S_{it}({\bf r})$ are the electron spin operators
for the localized and itinerant components of the electron system,
respectively. The exchange coupling $J=N(0)\;F_0^a$ can be obtained
from the expression for the energy in the Landau Fermi liquid
phenomenology.

This interaction leads to a high temperature scattering time for the 
itinerant fluid given by 
\bqa 
\frac{\hbar}{\tau} &=& \frac{\pi^2}{2} N(0) J^2 S(S+1)\; n_L 
\nonumber 
\\ 
&=& \frac{\pi^2}{2 N(0)} (F_0^a)^2 S(S+1)\; n_L 
\eqa 
for scattering off the spin $S=1/2$ ferromagnetic component. This 
gives a resistance 
\beq 
 \rho_S = \frac{m^\ast}{2 n_p\, e^2\, \tau} = \pi^2 \frac{h}{4 g_v
 e^2}\, \frac{n_L}{2 n_p} \;({F_0^a})^2\; S(S+1).
\enq
For a narrow coexistence region, the Landau parameter $F_0^a$ should
be close to the Pomeranchuk critical value of $-1$; hence, the
conductance is solely determined by the ratio $x= n_L/n$:
\beq 
\rho_S\approx 
\frac{3\pi^2}{4g_v} \frac{h}{4 e^2} \frac{x}{1-x}. 
\enq 
Notice that this result implies that there can be a large variation of 
the resistivity in the critical region of the phase diagram. The scale 
for resistance is the prefactor  
\beq 
\rho^\ast_S\approx 
\frac{3 \pi^2}{4g_v} \frac{h}{4 e^2}, 
\enq 
and if near the MIT transition region the fractions of the two
components are close, the ratio $x/(1-x)$ should be of order
unit. Hence, the high temperature resistance near the ``separatrix''
line should be of order $\rho^*_S$, which gives a conductivity of
order $\sigma \approx 0.3 \, \sigma_0$. This simple argument
highlights the importance of considering the electron-electron
interactions when calculating the resistivity of the system near the
ferromagnetic instability ($F_0^a=-1$).

  
\section{Landau Free Energy -- Mean-Field Phase Diagram} 
\label{sec:one}
 
In this section we discuss the transition between ferromagnetism and
p-wave superconductivity. In the context of Ruthenates it has been
shown that the ferromagnetic to superconductor transition can be
described in terms of an SO(10) model\cite{sigrist,obs0} Here we do
not take the high symmetry approach; instead, we simply write a
Landau-Ginzburg free energy that combines both p-wave
superconductivity and ferromagnetism (we assume the system to be
homogeneous):\cite{leggett}
\bqa 
\label{eq:freeenergy}  
F[{\bf m}, \psi, {\bf d}, {\bf d}^\ast] & = & \frac{b\, m^2}{2} + 
\frac{m^4}{4} + a\, \psi^2 + \delta\, \psi^2 m^2 \nonumber \\ & & 
+ \frac{\psi^4}{2} \int \frac{d\theta}{2\pi} \left\{ |{\bf 
d}(\theta)|^4 + \left[ i{\bf d}(\theta) \times {\bf d}^\ast(\theta) 
\right]^2 \right\} \nonumber \\ & & - \gamma\, \psi^2 \int 
\frac{d\theta}{2\pi} \left[ i{\bf d}(\theta) \times {\bf 
d}^\ast(\theta) \right] \cdot {\bf m}, 
\eqa 
where ${\bf m}$ is the ferromagnetic order parameter and $\psi$ and
${\bf d}(\theta)$ are amplitude and vector parts used to describe the
p-wave pairing order parameter at the Fermi surface:
\bqa
\Psi_{\alpha\beta} (\theta) = i\psi \sum_{k=1}^3 [\sigma_k
\sigma_2]_{\alpha\beta} d_k(\theta)
\label{true-order-parameter} \, ,
\eqa
with the 3D vector ${\bf d}(\theta)$ obeying the normalization
condition
\beq 
\int \frac{d\theta}{2\pi} |{\bf d}(\theta)|^2 = 1. 
\enq 
Here $\sigma_k$ with $k=1,2,3$ are Pauli matrices. 
We assume that the coefficients $\delta$ and $\gamma$ appearing in the
free energy are nearly independent of electron density $n$, while $a =
\alpha (n-n_A)$ and $b = \beta (n-n_B)$. All $\alpha$, $\beta$,
$\gamma$, and $\delta$ are positive.
 
Clearly not all fourth order terms allowed by symmetry have been taken
into account in Eq. (\ref{eq:freeenergy}). The inclusion of all terms
would render the analysis extremely difficult even at the mean-field
level. Thus, the choice manifest in Eq. (\ref{eq:freeenergy}) should
be considered as the simplest one that reproduces the phases discussed
in previous sections.
 
It is worth noticing that the expectation value of the Cooper pair 
total spin operator at a point $\theta$ of the Fermi surface is given 
by  
\beq \langle \hat{\bf S} \rangle = i\psi^2\, {\bf d}(\theta) \times 
{\bf d}^\ast(\theta). 
\enq  
Thus when ${\bf d}(\theta)$ is a real vector, apart from a overall
phase factor, $\langle \hat{\bf S} \rangle$ = 0. In this case, named
unitary, one can show that ${\bf d}(\theta)$ defines a direction along
which the spin operator $\hat{\bf S}$ has eigenvalue zero.
 
One could in principle minimize the free energy with respect to all
ten real parameters, namely, $\psi$, $m_i$, $d_i$, $d_i^\ast$, with
$i=1,2,3$, using a Lagrange multiplier to enforce the normalization
condition. Instead, we restrict the form of the vector ${\bf
d}(\theta)$ to certain classes, following the treatment used for
$^3$He.\cite{leggett} For instance, the analogous to the B phase or
Balian-Werthamer (BW) phase of $^3$He would correspond to the
isotropic (nodeless) choice
\beq 
{\bf d}(\theta) = \hat{e}_1\, \cos\theta + \hat{e}_2\, \sin\theta, 
\enq 
which is clearly unitary. Another possible unitary choice (also
nodeless), resembling the A phase or Anderson-Brinkman-Morel (ABM)
phase of $^3$He is
\beq 
{\bf d}(\theta) = \hat{e}_3 ( \cos\theta \pm i\, \sin\theta). 
\enq 
(The latter case seems to be relevant to the superconducting phase of
Sr$_2$RuO$_4$.\cite{manfred,baskaran}) Here, we use neither
choice, but adopt instead a parameterization that allows for a
nonunitary ${\bf d}(\theta)$, thus $\langle {\bf S} \rangle \neq
0$. Recall that there exists a ferromagnetic coupling between the
local magnetization and the Cooper pair total spin (the last term in
the free energy). A unitary order parameter would make this coupling
to vanish identically. Our choice will be
\beq 
\label{eq:parametrization} 
{\bf d}(\theta) = \frac{z}{\sqrt{2}} \left( w\, \hat{e}_1 + 
w^\ast\, \hat{e}_2 \right), 
\enq 
where $z \equiv e^{i\theta}$ and $|w|^2 = 1$. One may ask how this
comes about. The reasoning is simple. First, we look for a minimum
energy configuration and therefore fix it to be nodeless. Second, for
a 2D p-wave paired state, we must have $d_i(\hat{n}) = D_{i\alpha}
n_\alpha$, with $i=1,2,3$, and $\hat{n} = (\cos\theta,
\sin\theta)$. Thus, we can write that ${\bf d}(\theta) = z\, {\bf v}_1
+ z^\ast\, {\bf v}_2$, with ${\bf v}_1$ and ${\bf v}_2$ two complex
three-dimensional vectors that do not depend on $\theta$, but must
obey the relation $|{\bf v}_1|^2 + |{\bf v}_2|^2 = 1$ due to the
normalization condition on ${\bf d}(\theta)$. It is straightforward to
check that the BW and ABM states correspond to the choices ${\bf
v}_{1,2} = (\hat{e}_1 \mp i\hat{e}_2)/\sqrt{2}$ and ${\bf v}_1 =
\hat{e}_3$, ${\bf v}_2 = 0$, respectively. On the other hand, our
choice of parameterization can be obtained by setting, for instance,
${\bf v}_1 = (w\, \hat{e}_1 + w^\ast \hat{e}_2)/\sqrt{2}$ and ${\bf
v}_2 = 0$. The choice is not unique and has to be considered as an
{\it Ansatz}.
 
We can now proceed with the minimization of the free energy. With our 
choice for ${\bf d}(\theta)$, we find that 
\beq 
|{\bf d}(\theta)|^2 = 1 
\enq 
identically and 
\beq 
i\, {\bf d}(\theta) \times {\bf d}^\ast(\theta) = -\sin \phi\, {\hat 
e}_3, 
\enq 
where $w \equiv e^{i\phi/2}$. As a result, 
\bqa 
F[\psi^2,{\bf m},\phi] & = & a\, \psi^2 + \frac{\psi^4}{2} \left[ 1 + 
\sin^2 \phi \right] + \frac{b\, m^2}{2} + \frac{m^4}{4} \nonumber \\ & 
& + \delta\, \psi^2 m^2 + \gamma\, \psi^2 m_3\, 
\sin \phi. 
\eqa 
(Notice that we do not need to use a Lagrange multiplier to enforce 
the normalization condition anymore.) The minimization equations read 
\bqa 
\label{eq:psi} 
0 & = & a + \psi^2 [1 + \sin^2 \phi] + \delta\, m^2 + \gamma \, m_3\, 
\sin \phi, \\ 
\label{eq:m12} 
0 & = & m_{1,2} \left( b + m^2 + 2\delta\, \psi^2 \right), \\ 
\label{eq:m3} 
0 & = & m_3 \left( b + m^2 + 2\delta\, \psi^2 \right) + \gamma\, 
\psi^2\, \sin \phi, \\ 
\label{eq:phi} 
0 & = & \psi^2\, \cos \phi \left[ \psi^2\, \sin \phi + \gamma\, m_3 
\right]. 
\eqa 
After Eq. (\ref{eq:m12}), we can set $m_{1,2} = 0$ without loss of
generality (thus $m = m_3$ hereafter). Equation (\ref{eq:phi}),
however, is satisfied only when $\psi^2 =0$, or $\cos \phi = 0$, or $m
= -\psi^2/\gamma \, \sin \phi$. In our analysis of the possible
solutions to these equations we will be looking for the following
sequence of phases as the electron density is lowered: metal/p-wave
pairing/p-wave pairing + ferromagnet/ferromagnet.

We should stress that because of the dimensionality and the symmetry of
the order parameter we always have
\bqa
\langle \Psi_{\alpha,\beta} \rangle = 0
\label{no-order}
\eqa
at any finite temperature, although $\psi \neq 0$ and $\langle {\bf S}
\rangle \neq 0$. This is possible because, on average, $\langle {\bf
d}(\theta) \rangle = 0$ and therefore Eq. (\ref{no-order}) follows
directly from Eq. (\ref{true-order-parameter}).

\subsection{P-wave Paired Phase} 

Let us assume that $n_A \gg n_B$. Thus, starting from a paramagnetic
phase and lowering the density, we first encounter a second order
phase transition to a paired state ($\psi \neq 0$) at $n=n_A$. Indeed,
setting $m = 0$, we find from Eqs. (\ref{eq:m3}) and (\ref{eq:phi})
that $\sin \phi = 0$ and
\beq 
\psi^2 = -a = \alpha (n_A - n). 
\enq 
Presumably, in Si-MOSFETs, $n_A$ is a high density, outside the range
explored in the experiments that probe the MIT. Notice that since
$\sin\phi=0$, $\langle \hat{\bf S} \rangle = 0$ and this p-wave paired
phase is unitary.

\subsection{P-wave Paired + Ferromagnetic Phase Coexistence} 

Since $n_B \ll n_A$, we expect ferromagnetism to appear at much lower
densities ($n_B$ should close to the critical density of the MIT). In
this range, we may simple take $a \approx -\alpha n_A$. Looking for a
solution with $\psi^2 \neq 0$, $m \neq 0$, and $\sin \phi
\neq 0$ simultaneously, we obtain from Eq. (\ref{eq:phi}) that
\beq
\sin \phi = -\gamma\, \frac{m}{\psi^2}.
\enq
Now $\langle \hat{\bf S} \rangle = \gamma m\, \hat{e}_3 \neq 0$ and
consequently the pairing is nonunitary. The Cooper pair spin points
along the direction of spontaneous ferromagnetization. Solving the
other two equations for $m$ and $\psi^2$, we find
\beq
\label{eq:mag} 
m^2 = \frac{\beta(n_0 - n)}{1 - 2\delta^2}  
\enq 
and 
\beq 
\label{eq:super} 
\psi^2 = \frac{\delta\beta(n - n_1)}{1 - 2\delta^2},
\enq
respectively, with
\beq
n_0 = n_B + \frac{\gamma^2}{\beta} - \frac{2\alpha\delta\, n_A}{\beta}.
\enq
and
\beq
n_1 = n_B + \frac{\gamma^2}{\beta} - \frac{\alpha\, n_A}{\delta\beta}.
\enq
Notice that we need $\delta < 1/\sqrt{2}$ for stability. It is simple
to see that $m$ and $\psi$ go continuously to $0$ and $-\alpha\, n_A$,
respectively, as $n\rightarrow n_0$. The transition is again of second
order and $n_0$ represents the critical density where the
ferromagnetism sets. The pairing order parameter $\psi$ decreases as
one crosses $n_0$.

\subsection{Full Cooper Pair Spin Polarization} 
 
Once ferromagnetism appears, the Cooper pair total spin tends to align
itself with the ferromagnetic order parameter, ${\bf m}$. Full
polarization of the pair spin, $\langle \hat{\bf S} \rangle = 1 \cdot
\hat{e}_3$, occurs when $\sin \phi = -1$ and the onset is marked by
the point where $\psi^2 = \gamma\, m$. Combining this relation with
Eqs. (\ref{eq:mag}) and (\ref{eq:super}) we can find at which electron
density one reaches full spin polarization by solving the following
equation for $n_{\rm pol}$
\beq 
(n_{\rm pol} - n_1) = \frac{\gamma}{\delta} \sqrt{\frac{1-2\delta^2}
{\beta}} \sqrt{n_0 - n_1}.
\enq 
It is clear that the order $n_1 < n_{\rm pol} < n_0$ is obeyed,
constraining full pair spin polarization to occur while $\psi$ is
still finite. For electron densities lower than $n_{\rm pol}$, the
pairing order parameter continues to drop, while the ferromagnetic
order parameter grows. It is important to remark that although all
Cooper pairs are spin polarized, not {\it all} electrons in the system
are spin polarized. As discussed in Section \ref{sec:two-fluid}, there
also exist unpaired electrons in the whole range $n_c < n<n_0$. In the
coexistence region $n_c < n < n_{\rm pol}$, Eqs. (\ref{eq:psi}) and
(\ref{eq:m3}) are reduced to
\beq 
\label{eq:condic1} 
a + 2 \psi^2 + \delta\, m^2 - \gamma\, m = 0 
\enq 
and 
\beq 
\label{eq:condic2} 
m(b + m^2 + 2\delta\, \psi^2) - \gamma\, \psi^2 = 0. 
\enq 
%

\subsection{Ferromagnetism} 

As the density is lowered further than $n_{\rm pol}$, the p-wave
pairing parameter drops and the magnetization (connected to the spin
polarization of unpaired electrons) grows. At the point where the
pairing vanishes, Eq. (\ref{eq:condic2}) yields $m^2 = -b$. The
critical density $n_c$ can then be found by solving
Eq. (\ref{eq:condic1}), namely,
\beq 
 \left( n_B - \frac{\alpha\, n_A}{\delta \beta} \right) =
 \frac{\gamma} {\delta\sqrt{\beta}} \sqrt{(n_B - n_c)}.
\enq 
For $n<n_c$, we have $m = \sqrt{-b}$ and the magnetization may
increase up to its limit value. Obviously, this mean-field treatment
does not take into account quantum fluctuations or the interplay
between spin interactions and the Anderson localization. Whether these
effects happen above or below $n_c$ will depend on the microscopic
details which are outside the scope of the Landau-Ginzburg
phenomenology.
 
The exact form of the mean-field phase diagram resulting from the
minimization of the free energy depends on the values of the
coefficients $\alpha$, $\beta$, $\delta$, and $\gamma$. Besides the
constraint that all coefficients should be positive, it is necessary
the inequality $n_B > \alpha\, n_A/\delta\beta$ to hold in order to
$n_c$ be positive (more specifically, $n_1 < n_c < n_{\rm pol}$ under
this condition). This case is illustrate in Fig. \ref{phase diagram}.
Moreover, if $\gamma^2 \approx 2\alpha\delta\, n_A$, and both $\delta$
and $\alpha$ are sufficiently small, it is possible to have the
(experimentally accessible) critical densities $n_c$, $n_{\rm pol}$,
and $n_0$ very close to each other.

\section{Comparison with Experiments}  
\label{sec:experiments}  
 
In this section we analyze some of the recent experiments that probe 
the magnetic field dependence of the conductivity, and connect them to 
the theory we develop in this paper. Let us look separately into the 
parallel and perpendicular field experiments.

\subsection{Parallel Magnetic Field: Magnetotransport}  
  
Recent experiments have probed the dependence of the conductance in
Si-MOSFETs as a function of an in-plane magnetic
field.\cite{Shashkin,Vitkalov,Pudalov1} The experiments show that the
conductance saturates beyond a field $H_{\rm sat}$. The saturation
value is interpreted as the field needed to fully spin polarize the 2D
electron system. Further evidence for full spin polarization is
provided by an exact doubling of the period of the Shubnikov-de Haas
oscillations when the experiment is done in a slightly tilted magnetic
field and the in-plane component exceeds $H_{\rm
sat}$.\cite{vitkalov2,Okamoto}
 
In the experiments of Ref. \onlinecite{Shashkin}, it was observed that
the magnetic field $H_{\rm sat}$ ($B_c$ in the notation of
Ref. \onlinecite{Shashkin}) needed to saturate the magnetoconductance
at low temperatures scaled linearly with the difference $n-n_c$ over a
wide range of densities. Close to saturation, the scaled
magnetoconductance curves taken at different densities followed the
same function of the ratio between $H/H_{\rm sat}(n)$.

In Ref. \onlinecite{Vitkalov}, the magnetoconductance was also found 
to saturate beyond a value of magnetic field $H_{\rm sat}$ that 
depended on the density. Near the MIT, the conductivity followed the 
scaling 
\beq  
\sigma(H,T,n)-\sigma_{\rm sat}(n) = f\left( \frac{H}  
{H_{\rm sat}(n,T)} \right). 
\enq  
Again, similarly to Ref. \onlinecite{Shashkin}, $H_{\rm sat} \propto 
n-n_0$ for a wide range of densities at low temperatures. However, 
very close to the transition, the density and temperature dependent 
saturation field $H_{\rm sat}$ extracted from the data behaved 
approximately as 
\beq  
H_{\rm sat}(n,T)=A(n)\sqrt{\Delta(n)^2+T^2}.  
\label{eq:Hsatexp}  
\enq  
The parameter $A(n)$ is weakly dependent on density, being almost 
constant for a range of densities, and increasing by about 20\% near 
the critical density $n_0$. The parameter $\Delta(n)$ is fitted to a 
form 
\beq  
\Delta(n)=\Delta_0\; (n-n_0)^\gamma,  
\label{eq:Deltaexp}  
\enq 
with $\Delta_0\approx 2.27$ and $\gamma\approx 0.6$.  
  
Let us now explain some of these experimental results using the
theoretical framework we propose in this paper. We argued in Section
\ref{sec:Landau} that, despite the large interactions, we can still
apply the Landau Fermi liquid phenomenology in order to understand
both the instability towards a ferromagnetic state and its precursor
p-wave paired state. One of the Landau Fermi liquid parameters,
$F_0^a$, crosses the Pomeranchuk's stability boundary, $F_0^a=-1$,
leading to the ferromagnetic instability of the 2D electron
system. Near the critical density $n_0$, we performed a Landau
expansion for the Landau Fermi liquid parameter as in Eq. (\ref{landau
expansion}). In Ref. \onlinecite{Shashkin} the degree of polarization
$\xi \equiv g^\ast \mu_B H_\parallel /2E_F$, where
\beq
g^\ast = \frac{g_0}{1+F_0^a}
\label{eq:geffective}
\enq
is the effective Land\'e $g$-factor renormalized by the Landau Fermi
liquid parameter $F_0^a$.\cite{obs1} Thus,
\beq  
\xi = \frac{1}{1+F_0^a}\frac{g_0 \mu_B H_\parallel}{2E_F}.
\label{eq:xi_linear}
\enq  
When the field $H_\parallel$ is sufficiently strong to fully spin
polarize the system, we have $\xi=1$. If we neglect nonlinear terms in
the susceptibility (which in principle could be important when the
spin polarization is large), the saturation field in the metallic
phase can be estimated from the linear response expression of
Eq. (\ref{eq:xi_linear}),
\beq  
H_{\rm sat}=\frac{2E_F}{g_0 \mu_B}\; (1+F_0^a) = \frac{2\pi
\hbar^2}{g_0 m^\ast \mu_B}\, \alpha_\parallel (n-n_0),
\label{eq:parallelHsat}  
\enq  
where we have used Eq. (\ref{landau expansion}) and $g_v = 2$ for the
conduction band in (100) Si-MOSFET (we have also introduced
$\alpha_\parallel$ to indicate that this is the value of $\alpha$ when
the magnetic field is parallel to the 2D electron gas). We assume that
the only Landau parameter that is crossing an instability is $F_0^a$;
the mass ratio $m^*/m_b$, controlled by the singlet Landau parameter
$F_1^s$, is non critical.\cite{Pudalov1,obs2} Hence, the linear
dependence on the density difference $n-n_0$ of the saturation field
$H_{\rm sat}$ measured experimentally follows from the Landau
phenomenology above.
 
Using the experimental data for the $H_{\rm sat}$ dependence on
$n-n_0$ from Refs. \onlinecite{Shashkin,Vitkalov} and
\onlinecite{Pudalov} ($m^\ast \approx 1.5 m_b$), we find that the
parameter $\alpha_\parallel \approx 0.6$ in the expansion of the
$F_0^a$ as a function of $\delta=(n-n_0)/n_0$.
 
Although we can explain the linear dependence of $H_{\rm sat}$ {\it
vs.} $n-n_0$ observed for a range of densities in the experiments, we
would like to point out that we cannot easily explain the scaling
behavior closer to the critical point, as reported in
Ref. \onlinecite{Vitkalov}. One problem could be that near the MIT,
with the field vanishing, and with the uncertainty in the position of
the ferromagnetic transition (notice again that in our theory there
should be {\it two} quantum critical points), the exact form for the
dependence of the saturation field $H_{\rm sat}$ as a function of $T$
and $n$ may be harder to obtain. For example, this may be the cause
for the density dependent prefactor $A(n)$ in Eq. (\ref{eq:Hsatexp}).
  
A general feature of the scaling Eq. (\ref{eq:Hsatexp}) that we can
explain, however, is why the finite temperature corrections are
quadratic in $T$ for $T\ll \Delta$ and linear in $T$ for $T\gg
\Delta$. The low-$T$ behavior follows simply from the finite-temperature  
Fermi liquid susceptibility. The high temperature behavior, on the 
other hand, is due to the fact that the susceptibility should, in this 
temperature regime, obey a Curie law.

\subsection{Perpendicular Magnetic Field: Quantum Hall Effect} 
                
While a parallel magnetic field couples only to the electrons via the
Zeeman coupling, a perpendicular magnetic field, $H_{\perp}$, also
couples to the orbital motion and produces Landau levels. In
high-density Si-MOSFET samples, far from the critical value $n_c$, and
for small perpendicular fields, it is observed that the electronic
states are localized. With increasing magnetic field the electrons
tend to delocalize because of the Lorentz force, leading to the
so-called floating of the extended states.\cite{laughlin} For
sufficiently large $H_{\perp}$, orbital effects start to play a role,
and a series of quantum Hall phases are observed in the
Si-MOSFETs.\cite{Kravchenko-qhe} Integer quantum Hall plateaus are
observed when an integer number of electrons becomes commensurate with
the number of flux quanta $\phi_0 = c h/e$ piercing the system,
\bqa 
n = \frac{H_{\perp}}{\phi_0} \nu, 
\label{landau fan} 
\eqa 
where $\nu$ is the magnetic filling factor (number of electrons per
flux quanta). In the case of high-density Si-MOSFET the bare Zeeman
splitting is very small compared with the cyclotron energy,
\bqa 
\hbar \omega_c = \frac{\hbar e H_{\perp}}{m_b c} \, , 
\label{omegac} 
\eqa 
implying that up and down states are nearly degenerate. Moreover, the
band structure dispersion contains two valleys that are almost
degenerate. Equation (\ref{landau fan}) for different integers $\nu$
gives the ``Landau fan'' in the $H_{\perp} \times n$ plane. Whenever
condition Eq. (\ref{landau fan}) is obeyed, the system sits at the
middle of a Hall plateau, where the Hall resistivity, $\rho_{xy}$,
becomes quantized in units of $h/e^2$ ($\rho_{xy} = h/\nu e^2$) while
the longitudinal resistance, $\rho_{xx}$,
vanishes.\cite{Kravchenko-qhe} The fact that the quantum Hall effect
occurs implies that Landau levels are occupied by well-defined
quasiparticles. This gives extra support to our assumption that a
Fermi liquid description is correct at intermediate temperatures and
magnetic fields.
 
For a given filling fraction $\nu$ and at lower densities, close to
but higher than the zero field $n_c$ (see
Ref. \onlinecite{Kravchenko-qhe}), the quantized Hall states are
lost. This occurs at some critical density $n_c(\nu)$, at which the
states at the center of the Landau band localize, and the longitudinal
resistance diverges as $T\rightarrow 0$. It is known\cite{Iorio90}
that $n_c(\nu)$ varies substantially along the ``Landau fan''. In
fact, it has been shown by Kravchenko and collaborators that if the
electron system is spin polarized in the plane (by a parallel magnetic
field) and a perpendicular field is applied, the localization effects
occur already at higher densities.\cite{sakr} This demonstrates the
importance of spin interactions in the problem: the correlation
effects are reduced when fully spin polarization is achieved, making
the system closer to noninteracting and thus more sensitive to
localization effects. One of the most interesting observations is that
for $\nu=4,8,12$, corresponding to filling factors within the
cyclotron gaps, the localization occurs at higher densities (or
magnetic fields fields) than for $\nu = 2,6,10$, where filling factors
fall into spin gaps (recall that $g_v = 2$). This effect has not been
quantitatively explained so far. We show below that it can be
understood as the localization of the carriers due to the enhancement
of the spin susceptibility near the zero-field critical density $n_c$.
 
As pointed out earlier, at the localization transition there is an
enhancement of the magnetic susceptibility, and consequently of the
effective Land\'e $g$-factor, $g^\ast$. Therefore, the Zeeman
splitting energy
\bqa 
E_Z(n) = g^\ast \mu_B H_{\perp} 
\label{zeeman} 
\eqa 
becomes large and of order of the cyclotron energy. Here $g^\ast$
can be obtained from the Fermi liquid theory [see
Eq. (\ref{eq:geffective})],
\bqa
g^\ast = \frac{g_0}{\alpha_{\perp} \delta},
\label{geff}
\eqa
which diverges at the quantum critical point. This implies that there
is a level crossing between the down spin state of the $i$th Landau
level with the spin up state of the $i+1$th Landau level. When the
crossing occurs there is an excess magnetization in the system and, in
particular, at low field (around 1 Tesla), for the case of $\nu =4$
(see below), the system becomes fully polarized. The experimental
evidence is that at this point the carriers localize since the
longitudinal resistivity increases as $T\rightarrow
0$.\cite{Kravchenko-qhe} This observation implies that below the
critical density and in the presence of an applied magnetic field, the
localized state, independent of the direction of the field, is indeed
spin polarized.
 
To illustrate this effect, consider the situation when the magnetic
field is enough to produce a $\nu =4$ state. In this case there are
equal number of up and down spins filling the first Landau level with
energy $\hbar \omega_c/2$. At large densities the Zeeman energy is
insignificant when compared to $\hbar \omega_c$ and the up and down
spin states can be considered as degenerate. As the density is
decreased along the curve defined by Eq. (\ref{landau fan}) the
effective $g$ factor increases according to the susceptibility in
Eq. (\ref{divergences}). The first Landau level for up ($\uparrow$)
and down ($\downarrow$) spin states changes as a function of density
as
\bqa 
E_{i=1,\uparrow}(n) &=& \frac{1}{2} \hbar \omega_c - \frac{1}{2}
E_Z(n)
\nonumber 
\\  
E_{i=1,\downarrow}(n) &=& \frac{1}{2} \hbar \omega_c + \frac{1}{2}
E_Z(n),
\label{first} 
\eqa 
while the second Landau level changes its energy with respect to the
density as
\bqa 
E_{i=2,\uparrow}(n) &=& \frac{3}{2} \hbar \omega_c - \frac{1}{2}
E_Z(n)
\nonumber 
\\  
E_{i=2,\downarrow}(n) &=& \frac{3}{2} \hbar \omega_c + \frac{1}{2}
E_Z(n) \, .
\label{second} 
\eqa 
Thus, there is a critical density $n^\ast$ such that the previously
empty second Landau level for spin up, $E_{i=2,\uparrow}(n^\ast)$,
becomes degenerate with the first Landau level with spin down,
$E_{i=1,\downarrow}(n^\ast)$. Thus, $n^{\ast}$ is given by
\beq 
E_Z(n^\ast) = \hbar \omega_c \, .
\enq 
Using (\ref{zeeman}), (\ref{geff}) and $\mu_B = e \hbar/(2 m_0 c)$ we
find
\beq 
n^\ast \approx n_0 \left(1 + \frac{g_0}{2\alpha_\perp}
\frac{m_b}{m_0}\right) \, .
\enq 
Using $g_0 = 2$, $m_b = 0.2 m_0$, $n_0 = 0.8 \times 10^{11}$
cm$^{-2}$, and $n^\ast = 1.0 \times 10^{11}$ cm$^{-2}$
(Ref. \onlinecite{sakr}), we find $\alpha_\perp \approx 0.8$. This
value should be compared with the value of $\alpha_\parallel \approx
0.6$ found in the case of a parallel magnetic field. The agreement is
good and gives extra support to the idea that the localized state is
indeed ferromagnetic even when the field is perpendicular to the 2D
electron gas. We remark, however, that we do not expect the same
estimate of $\alpha_\perp$ to be applicable to the states $\nu=8$ and
$\nu=12$, since they vanish at densities sufficiently far from $n_c$
to invalidate the use of expansion implicit in Eq. (\ref{geff}).

\subsection{Classical Hall Effect}

A unusual experimental fact related to the behavior of the classical
Hall coefficient in Si-MOSFET's can also be accounted for by our
theory. It has been observed that the Hall resistance in Si-MOSFET at
low temperatures and at densities $n>n_c$ is insensitive to parallel
magnetic fields ranging from zero to $H_\parallel > H_{\rm
sat}$.\cite{vitkalov_hall} That is, there seems to exist a {\it
single} charge carrier component in the metallic phase, for all values
of $H_{\rm sat}$, instead of two independent spin up and down
components. This is consistent with the idea that the conducting fluid
present in the metallic phase is formed by {\it electron pairs}
(bosons), instead of up/down spin unpaired electrons
(fermions). Although the number of pairs decreases with an applied
parallel field (and consequently the number of localized electrons
increases, see Section \ref{sec:two-fluid}), notice that the
experiment is performed at {\it fixed} current.\cite{obs3}

\subsection{A New Experiment: Shot Noise}

The main prediction of our theory is that in the metallic phase, at
very low temperatures, the system is composed by incoherent p-wave
pairs of electrons. The bosons are incoherent because there is no long
range order in the system [see Eq. (\ref{no-order})] and therefore no
gap in the spectrum. This situation is very similar to the case of a
metallic bosonic liquid.\cite{Phillips2} Due to the lack of phase
coherence, the usual methods to measure the pair charge, such as the
Josephson effect, cannot be used. Instead, the simplest way to measure
the pair charge $2e$ is by making a constriction in the 2D electron
density profile via external gates and measure the shot noise on the
current across the constriction (for shot noise we must have $k_B T
\ll e V$ where $V$ is the voltage applied across the constriction).
Since the pairs behave as independent bosonic entities, the current
fluctuations should be quantized in units of the elementary charge
$2e$.\cite{shot-noise}

\section{Conclusions} 
\label{sec:conclusions} 
 
In this paper we argue that one can account, on the basis of Landau
Fermi liquid theory, for the recent experimental observations that the
characteristic magnetic field $H_{\rm sat}$ needed to saturate the
conductance in 2D Si-MOSFETs at low temperatures vanishes at a
critical value of the electronic density $n_0$. We propose a
phenomenological expansion for the Landau parameter
$F_0^a=-1+\alpha(n-n_0)/n_0$ in terms of the electronic density $n$
that drives the system ferromagnetic, through a Pomeranchuk
instability, at $n_0$. As one approaches the instability, the spin
susceptibility is greatly enhanced, requiring smaller magnetic fields
to fully spin polarize the system. At the critical point an
arbitrarily small magnetic field fully polarizes the system, since the
susceptibility (at zero temperature) diverges.

We also analyze the effects of a perpendicular magnetic field through
the system, in the quantized Hall regime, and show that the critical
density for the localization of the $\nu=4$ state as compared to the
$\nu=2$ state can be accounted for by considering the crossover
between a spin polarized and unpolarized state due to the enhancement
of the Land\'e $g$ factor. The values of the parameter $\alpha$
estimated separately from parallel and perpendicular field experiments
agree within 20\%.
 
In the paramagnetic side, but close to the instability, the enhanced
spin fluctuations can lead to an attractive interaction in the spin
triplet channel, similarly to superfluid $^3$He. We analyze a
Landau-Ginzburg mean-field theory that combines p-wave
superconductivity and ferromagnetism, and find two quantum critical
points as a function of density, $n_0$ where ferromagnetism begins,
and $n_c$ where p-wave pairing ceases. There is an intermediate range
of densities where p-wave pairing and ferromagnetism coexist. In this
range, the p-wave state is in a nonunitary phase.
 
For large enough densities, above a value $n_A$, the paramagnon exchange
mechanism responsible for attractive interactions should cease (when, for
example, $F_0^a>0$). At these densities, an Anderson insulating state,
similarly to the case of noninteracting electrons, should occur. However, the
presence of the p-wave paired state for $n_c<n<n_A$ does not rule out the
possibility of a conducting phase. Since the order parameter for the p-wave
state is a vector, no order (even algebraic) exists at finite temperature,
and true superconductivity should only occur at $T=0$. In this paper we do
not present any explanation why the p-wave pairing would lead to the
conducting phase at finite $T$; however, we strongly believe that if this
correlated state exists in the 2D electronic systems, it may provide the
origin of the extended state that continues to accumulate experimental
support.\cite{Sarachik}


Summarizing, we propose that the metallic state close to the
metal-insulator transition in the 2D electron gas problem is due to
the existence of a paired p-wave state close to a ferromagnetic
insulating phase. The pairing is generated by long wavelength magnetic
fluctuations close to the quantum critical point. We describe the
pairing within Landau's Fermi liquid phenomenology and show that it
provides a consistent description of the data for parallel and
perpendicular magnetic fields. Moreover, we propose new 
shot noise experiments that can test our theory.

 
\section{Acknowledgements} 
 
The authors would like to thank E. Demler, P. Lee, P. Phillips,
S. Sachdev, N. Sandler, and M. Sigrist for enlightening
discussions. We are particularly indebted to S. Kravchenko,
M. Sarachik, and S. Vitkalov for thoroughly discussing their
experimental data with us, and to S. Chakravarty for careful reading
of the manuscript and very stimulating discussions. Support was
provided by the NSF Grant DMR-98-76208, the Alfred P. Sloan Foundation
(C.C.), and the Brazilian agencies CNPq, FAPERJ, and PRONEX
(E.R.M.). A.H.C.N acknowledges partial support provided by a
Collaborative University of California - Los Alamos (CULAR) research
grant under the auspices of the US Department of Energy.


\end{multicols}  
 

\begin{references}
\bibitem{footKT} If the order parameter vector direction is pinned
at the 2DEG interfaces, the symmetry is reduced to $U(1)$ and there is
a KT transition at finite temperature; we would like to thank
S. Chakravarty for pointing this possibility to us.
\bibitem{obs2} Experiments confirm that singlet Landau parameters do
not behave critically: In Ref. \onlinecite{weizmann} it was found that
the compressibility (renormalized by $F_0^s$ in the Fermi liquid
theory) is a smooth function at the MIT.
\end{references}
\end{document}